\documentclass[pdflatex,sn-mathphys-num]{sn-jnl}

\usepackage{natbib}
\usepackage{graphicx}%
\usepackage{multirow}%
\usepackage{amsmath,amssymb,amsfonts}%
\usepackage{amsthm}%
\usepackage{mathrsfs}%
\usepackage[title]{appendix}%
\usepackage{xcolor}%
\usepackage{textcomp}%
\usepackage{manyfoot}%
\usepackage{booktabs}%
\usepackage{algorithm}%
\usepackage{algorithmicx}%
\usepackage{algpseudocode}%
\usepackage{listings}%

\theoremstyle{thmstyleone}%
\theoremstyle{thmstyletwo}%

\theoremstyle{thmstylethree}%

\begin{document}

\title[Shock acceleration in vortex driven magnetic fields of black holes]{Shock acceleration in vortex driven magnetic fields of black holes}


\author*[1,2]{\fnm{Z.N.} \sur{Osmanov}}\email{z.osmanov@freeuni.edu.ge}



\affil[1]{\orgdiv{School of Physics, Free University of Tbilisi, 0159 Tbilisi, Georgia} }

\affil[2]{\orgdiv{E. Kharadze Georgian National Astrophysical Observatory, Abastumani 0301, Georgia} }



\abstract{The aim of this work is to investigate particle acceleration by shock waves in the magnetospheres of supermassive black holes, where the magnetic field is vortex-driven. For this purpose, we investigated the first-order Fermi acceleration process at relativistic shocks in the magnetospheres of supermassive black holes. In our analysis, we considered synchrotron radiation and inverse Compton scattering as the principal energy-loss mechanisms limiting the maximum attainable particle energy. We investigated the acceleration of both protons and electrons and demonstrated that, for a typical supermassive black hole with a mass of $10^8$ solar masses, protons can be accelerated to energies of several hundred PeV, whereas electrons can attain energies of up to approximately 120 GeV.}




\maketitle

\section{Introduction}

One of the fundamental problems in astrophysics is understanding the origin of very-high-energy (VHE) cosmic rays \citep{uhecr}. According to recent studies, supermassive black holes and pulsars are considered among the leading candidate sources of VHE particles \citep{agn,neutr,pulsars}.

In recent years, significant progress has been made in the study of the magnetocentrifugal acceleration mechanism, which has been extensively investigated in compact astrophysical objects \citep{NO,review}.

One of the most efficient particle acceleration processes is the Fermi mechanism, together with its various modifications \citep{fermi,bell1,bell2}. It is well known that the efficiency of Fermi acceleration depends critically on the magnetic field strength. Recent studies indicate that the magnetic fields in the vicinity of black holes may be significantly stronger than those expected from standard astrophysical processes.

In particular, the study shows that an ultra-strong magnetic field, $B$, may be generated around rapidly rotating black holes by a powerful vortex \citep{vortex}. The similar effect can be generated by a theoretical possibility of a massive photon \citep{DOZ}. The theoretical upper limit of the field is reached when the magnetic field energy, $B^2r_g^3$ (we omit dimensionless numbers) becomes comparable to the total energy of the black hole, $Mc^2$, leading to the maximum $B$-value
\begin{equation} \label{BBound}
    B_m \simeq \frac{c^4}{MG^{3/2}}\simeq 2.4\times 10^{11}\times\frac{10^8M_{\odot}}{M}\; Gauss,
\end{equation}
where $r_g = GM/c^2$ and $M$ are the gravitational radius and mass of the black hole respectively, $G$ denotes the gravitational constant, $c$ is the speed of light, mass is normalized by the mass of typical supermassive black holes and $M_{\odot}$ denotes the solar mass. As is evident from Eq. (\ref{BBound}), the magnetic induction is much larger than the one derived by the conventional mechanisms. Using this magnetic field configuration, we studied the electron–positron pair production rate in the magnetospheres of supermassive black holes and showed that it is many orders of magnitude higher than the corresponding pair production rate in conventional magnetic field configurations \citep{pair1,pair2}.
 
In this work, we investigate the efficiency of type-I Fermi acceleration in the presence of vortex-induced magnetic fields and determine the maximum energies that particles can attain under such conditions.

The paper is organized as follows: 
in Section 2, we describe the acceleration mechanism and evaluate its efficiency for typical astrophysical black holes, and in Section 3, we summarize the main results obtained in this study.

\section{Main consideration}

Plasma flows in the vicinity of a black hole are well known to be supersonic, which generally leads to the formation of shock waves \citep{bland}.

Since the magnetic field is extremely strong, the Alfv\'en velocity is given by \citep{anile}
\begin{equation} \label{alfven}
\upsilon_A = c\left(\frac{B^2}{4\pi w+B^2}\right)^{1/2},
\end{equation}
where $w$ denotes the enthalpy density, which for non-relativistic cold plasma equals $nm_pc^2$, $n$ is the plasma number density and $m_p$ is the proton's mass. It is clear that for the considered magnetic field $nm_pc^2<<B$, therefore, the Alfv\'en velocity is close to the speed of light. Therefore, it is quite natural to assume that the relativistic propagation of shock waves constitutes a realistic physical mechanism.
 
We now estimate the maximum possible rate of energy gain
\begin{equation} \label{dEdt1}
\frac{dE}{dt}\simeq\frac{E}{t_{acc}},
\end{equation}
where $E$ is particles' energy and $t_{acc}$ represents the acceleration time-scale, which for sufficiently strong magnetic fields, can be approximated as to be $\eta r_L/c$ with $r_L\simeq E/eB$ representing the Larmor radius and $e$ - the proton's charge. $\eta\geq 1$ is a dimensionless parameter, representing an efficiency of the acceleration process. The larger the parameter, the slower the acceleration process. After combining Eq. (\ref{dEdt1}) with the expression of $t_{acc}$, one obtains the maximum possible energy gain in this shock acceleration process
\begin{equation} \label{dEdt2}
\frac{dE}{dt}\simeq\frac{1}{\eta}eBc.
\end{equation}

As is evident from this result, the acceleration process is highly efficient, and the particle energy is expected to increase rapidly. On the other hand, as the particle gains energy while moving through a strong magnetic field, it begins to emit synchrotron radiation. As a result, at sufficiently high energies, the energy losses due to synchrotron emission balance the energy gain, establishing an equilibrium beyond which the particle energy can no longer increase.

By taking the average synchrotron emission power into account,
\begin{equation} \label{Psyn}
P_{syn}\simeq\frac{4e^4B^2E^2}{9m^4c^7},
\end{equation}
by using a balance condition, $dE/dt\simeq P_{syn}$
one can estimate the maximum attainable energy of particles for the magnetic field $B = \alpha B_m$ ($\alpha\leq 1$)
\begin{equation} \label{Emax}
E_{max}\simeq\frac{3m^2c^4}{2e\sqrt{\eta\alpha eB}}\simeq\frac{590}{\sqrt\eta}\times\left(\frac{m}{m_p}\right)^2\times\left(\frac{M}{\alpha\; 10^8M_{\odot}}\right)^{1/2}\; TeV.
\end{equation}
Theoretical analysis shows that the minimum magnetic field strength can be as much as $10^6$ ($\alpha = 10^{-6}$) times smaller than the aforementioned theoretical upper limit \citep{DOZ}. In general, another mechanism that may potentially impose an upper limit on the attainable particle energy is inverse Compton scattering. However, as is well established in cosmic-ray physics, this process is strongly suppressed for protons \citep{ahar}.

\begin{figure}
  \centering {\includegraphics[width=10cm]{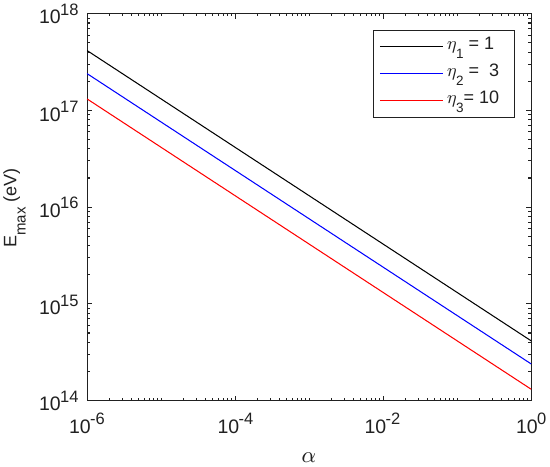}}
  \caption{Maximum attainable energy of protons versus $\alpha$. The set of parameters is: $m = m_p$, $M = 10^8M_{\odot}$.}\label{fig1}
\end{figure}

On Fig. 1 we plot $E_{max}$ versus the dimensionless parameter $\alpha$. The set of parameters is: $m = m_p$, $M = 10^8M_{\odot}$. As it is clear from the plots, for the aforementioned set of parameters, the shock acceleration might provide energies in the interval from $100$ TeV up to $400$ PeV.

For electrons, inverse Compton scattering must also be taken into account. As is well known, in the Thomson regime the corresponding cooling power is given by $4\sigma_TE^2U_{ph}/(3m_e^2c^4)$ \citep{rybicki} with the thermal energy density $U_{ph} = 4\sigma T^4/c$ \citep{carroll}, where $\sigma_T$ is the Thomson cross section, $\sigma$ represents the Stefan-Boltzmann constant and $T$ is the temperature of the accretion disk. The corresponding energy balance leads to the maximum energy of electrons 
\begin{equation} \label{EIC}
E_{IC}\simeq\frac{m_ec^3}{4T^2}\times\left(\frac{3e\alpha}{\eta\sigma_T\sigma}\right)^{1/2}.
\end{equation}

Taking into account the estimated accretion temperature \citep{carroll}
\begin{equation} \label{T}
T\simeq\left(\frac{GM\dot{M}}{8\pi\sigma R_{S}}\right)^{1/4},
\end{equation}
where $\dot{M} = \pi(GM/u_{\infty}^2)^2m_pn_{\infty}u_{\infty}$ is an accretion rate \citep{shapiro}, $u_{\infty}$ and $n_{\infty}$ represent the sound speed and the particle number density respectively, far from the black hole and $R_S = 2GM/c^2$ denotes the  Schwarzschild radius. One can straightforwardly show that by assuming a natural astrophysical parameter $n_{\infty} =1 cm^{-3}$, the corresponding energy violates a condition $E_{IC}kT/(m^2c^4)<<1$ ($k$ is the Boltzmann constant) when the Thomson regime works.

Therefore, inverse Compton scattering must be considered in the Klein–Nishina regime, where the corresponding cooling power is given by $P_{KN}\simeq\frac{\sigma (m_eckT)^2}{16\hbar^3}\times \left(ln\frac{4\gamma kT}{m_ec^2}-1.981\right)$ \citep{blum}. Then considering the cooling time-scale, $t_{KN}\simeq\gamma m_ec^2/P_{KN}$, one can see that it is by many orders of magnitude larger than the acceleration time-scale
\begin{equation}
\label{ratio} 
\frac{\tau_{KN}}{\tau_{acc}}\simeq\frac{4.3\times 10^{12}\times\alpha}{\eta\left(ln\frac{4\gamma kT}{m_ec^2}-1.981\right)}.
\end{equation}
The obtained results clearly indicate that inverse Compton losses, in neither the Thomson nor the Klein–Nishina regime, have a significant effect on the electron acceleration process. Therefore, as in the case of protons, the limiting factor that determines the maximum attainable energy is solely synchrotron radiation.

\begin{figure}
  \centering {\includegraphics[width=10cm]{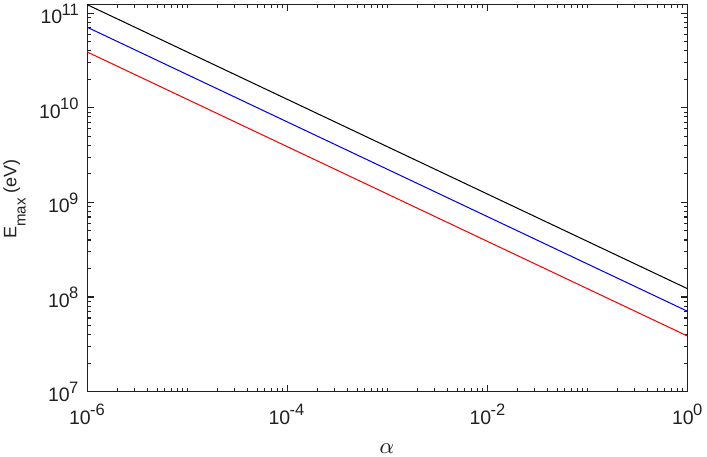}}
  \caption{Maximum attainable energy of electrons versus $\alpha$. The set of parameters is same as in the previous figure, except: $m = m_e$.}\label{fig2}
\end{figure}

On Fig. 2 we plot the electrons' energies versus $\alpha$. The set of parameters is the same as in the previous figure, except: $m = m_e$. As is evident from the figures, electrons can be accelerated to energies ranging from approximately 40 MeV to 120 GeV.

The behavior of the maximum energy shown in both figures is entirely expected. In particular, while a stronger magnetic field increases the acceleration efficiency, the synchrotron power scales as the square of the magnetic field strength. As a result, synchrotron losses increase more rapidly than the acceleration rate, making radiative losses the dominant limiting factor. Consequently, the maximum attainable energy, $E$, is a monotonically decreasing function of $\alpha$. That is why considering the conventional magnetic field, the shock acceleration leads to higher values of energies \citep{rieger}.

As emphasized above, the acceleration mechanism investigated in this work is strongly constrained by synchrotron losses. It is therefore worthwhile to explore alternative acceleration mechanisms in which this limiting factor plays a less significant role. A promising candidate is magnetocentrifugal acceleration \citep{NO}. In a subsequent study, we plan to investigate this particular mechanism in the presence of vortex-driven magnetic fields.

\section{Conclusions}
In this work, we investigated first-order Fermi acceleration at relativistic shocks in the magnetosphere of a supermassive black hole threaded by a vortex-driven magnetic field.

In our analysis of the acceleration process, we took into account the principal energy-loss mechanisms, namely synchrotron radiation and inverse Compton scattering. We demonstrated that for both electrons and protons, synchrotron radiation is the dominant cooling process and therefore constitutes the primary factor limiting the maximum attainable particle energy.

Our analysis showed that the proposed acceleration mechanism can accelerate protons to energies ranging from approximately $100$TeV to $400$PeV, while electrons can attain energies between approximately $40$MeV and $120$GeV.

Since our analysis has shown that the maximum attainable particle energies are strongly constrained by synchrotron losses, whose cooling rate depends sensitively on the magnetic field strength, our next objective is to investigate magnetocentrifugal acceleration, where the synchrotron emission does not impose any limitations on  acceleration processes.
\bmhead{Acknowledgements}

The research was supported by the Shota Rustaveli National Science Foundation of Georgia (SRNSFG) Grant: FR-24-1751.


\begin{itemize}
\item The research was supported by the Shota Rustaveli National Science Foundation of Georgia (SRNSFG) Grant: FR-24-1751.
\item No conflict of interests.
\item No datasets were generated or analysed during the current study.
\end{itemize}

\end{document}